\documentclass[%
 preprint,
 amsmath,amssymb,
 aps,
 onecolumn,
 superscriptaddress
]{revtex4-2}

\usepackage{graphicx}
\usepackage{dcolumn}
\usepackage{bm}

\usepackage[T1]{fontenc}
\usepackage[utf8]{inputenc}
\usepackage{color}
\usepackage{babel}
\usepackage{array}
\usepackage{amsmath}
\usepackage{amssymb}
\usepackage{graphicx}
\usepackage[pdfusetitle,
 bookmarks=true,bookmarksnumbered=false,bookmarksopen=false,
 breaklinks=false,pdfborder={0 0 1},backref=false,colorlinks=false]
 {hyperref}
\hypersetup{
 colorlinks=true, linkcolor=blue, urlcolor=blue, citecolor=blue}

\begin{document}

\title{A collider as a quantum computer}

\author{Wei Xie}
\email{xiewei@ctgu.edu.cn }

\affiliation{Department of Physics, Liaoning Normal University,
No. 850 Huanghe Road, Dalian 116029, China.}
\affiliation{\selectlanguage{american}%
College of Mathematics and Physics, China Three Gorges University,
No. 8 University Road, Yichang 443002, China.}
\affiliation{\selectlanguage{american}%
Center for Astronomy and Space Sciences, China Three Gorges University,
No. 8 University Road, Yichang 443002, China.}

\author{Ji-Chong Yang}
\email{yangjichong@lnnu.edu.cn (Corresponding author)}

\affiliation{Department of Physics, Liaoning Normal University, No. 850 Huanghe
Road, Dalian 116029, China.}
\affiliation{Center for Theoretical and Experimental High Energy Physics, Liaoning
Normal University, No. 850 Huanghe Road, Dalian 116029, China.}

\date{\today}

\begin{abstract}
Scattering processes in high-energy physics are inherently quantum mechanical, yet are typically analyzed at the level of final states, where entanglement appears as a property of the outcome rather than a consequence of the underlying dynamics. 
We reformulate scattering at the level of the process itself by representing helicity transition matrices as quantum circuits. 
Once the kinematic configuration and scattering channel are fixed, the problem reduces to a finite-dimensional quantum map, making a circuit description natural. 
Within this framework, an example of the process $e^+e^-\to \mu^+\mu^-$ is shown, which decomposes into unitary and non-unitary components, corresponding to coherent mixing and postselection effects. 
This representation reorganizes the amplitude into distinct operational elements, providing a perspective in which collider processes can be viewed as constrained quantum circuits and their entanglement structure can be understood in terms of the underlying circuit dynamics, opening the door to analyzing their properties using the language of quantum information.
\end{abstract}

\maketitle


\section{Introduction}

There have been rapid developments at the interface of high energy physics and quantum information in recent years~\citep{Braunstein:2005zz, Weedbrook:2011wxo, Lloyd:1998jk, Ferraro:2005hen, Pfister:2019apg, Wiese:2013uua, Banuls:2019bmf, Bergner:2024qjl, Illa:2024kmf, Jordan:2012xnu, Jordan:2011ci, Jordan:2014tma, Zohar:2021nyc, Martinez:2016yna, Klco:2018kyo, Kokail:2018eiw, Cobos:2025krn, Gaz:2025ptu, Ciavarella:2024fzw, Balaji:2025afl, Atas:2021ext, Cataldi:2025cyo, deForcrand:2009zkb}. In particular, a growing body of work has explored how entanglement is generated in particle collisions~\citep{Blasone:2024dud,Kraus:2000nly,Fonseca:2021uhd,Peschanski:2016hgk,Seki:2014cgq,Cervera-Lierta:2017tdt,Liu:2025frx,Liu:2025qfl,Kowalska:2025qmf,Aoude:2024xpx,McGinnis:2025brt,Low:2021ufv,Guo:2026yhz}, typically by analyzing the quantum correlations present in the final-state particles. Within this perspective, scattering amplitudes are used to construct final-state density matrices, from which various entanglement measures are extracted.

While this approach has led to many interesting observations, it treats the scattering process itself only implicitly, as a map that produces the final state. As a result, entanglement is viewed primarily as a property of the outcome, rather than as a consequence of a structured dynamical process. In particular, the organization of the underlying quantum evolution---including how different contributions combine and how coherence and selection effects enter---is not directly exposed in the final-state description.

In this work, we instead formulate scattering at the level of the process itself. Once a scattering channel and its kinematic configuration are fixed, the dynamics reduces to a finite-dimensional quantum map acting on internal degrees of freedom such as helicity. This observation makes it natural to represent the scattering process as a quantum circuit, in which the amplitude is interpreted as an operator acting between input and output states. The use of quantum circuits as a general framework for describing quantum processes is well established in quantum information theory~\cite{nielsen2002quantum,preskill1998lecture}.

This circuit-based perspective provides a different organization of the same physical information. Rather than encoding the dynamics in a set of coefficients that mix kinematic factors and couplings, the process is decomposed into operational elements, such as unitary transformations and non-unitary components associated with postselection. In this way, features that are implicit in the amplitude description become explicit in the circuit representation, allowing the structure of the process to be analyzed in terms of its constituent operations.

From this viewpoint, a collider experiment may be regarded as implementing a constrained quantum circuit, where the form of the circuit is determined by the interaction structure and its parameters are controlled by kinematics. This reformulation opens a complementary route to studying scattering processes, in which questions about entanglement can be addressed in terms of the properties of the underlying quantum process rather than solely through the analysis of final states.

In the following, we illustrate this approach using a simple lepton scattering process. We construct the corresponding quantum circuit representation of the helicity transition matrix and analyze its structure, demonstrating how the circuit language reorganizes the information contained in the amplitude.

\section{\protect\label{sec:The-model}The model}

\subsection{\label{sec:IIA}Scattering as a finite-dimensional linear transformation}

We begin with the standard description of a scattering process in terms of the $S$-matrix. 
For a fixed scattering channel and given kinematic configuration, the transition amplitude between an initial state $|\alpha\rangle$ and a final state $|\beta\rangle$ is encoded in the $T$-matrix element,
\begin{equation}
\langle \beta | T | \alpha \rangle,
\end{equation}
where $S = \mathbf{1} + i T$. 

In practical calculations, one typically works in a basis of definite quantum numbers for the external particles. 
For processes involving fermions, a convenient choice is the helicity basis, in which both initial and final states are labeled by discrete helicity indices. 
Once the particle content and kinematic configuration are fixed, the set of allowed helicity configurations is finite, and the $T$-matrix can be represented as a finite-dimensional matrix acting on this space.

From this point of view, the scattering amplitude defines a linear relation between the space of initial helicity states and that of final helicity states. That is, for a given process, one may regard $T$ as a linear operator
\begin{equation}
T: \mathcal{H}_{\mathrm{in}} \rightarrow \mathcal{H}_{\mathrm{out}},
\end{equation}
where $\mathcal{H}_{\mathrm{in}}$ and $\mathcal{H}_{\mathrm{out}}$ are finite-dimensional Hilbert spaces spanned by the corresponding helicity bases. All dynamical information of the process, including the dependence on couplings and kinematic variables, is contained in this operator.
One may intuitively view this as selecting a sub-block of the full $S$-matrix associated with a given scattering channel.
More generally, this corresponds to restricting the full unitary evolution to a reduced space of helicity states.
Since this space is not closed under the full dynamics, probability can flow to other sectors, and the resulting transformation is therefore not unitary.

It is worth emphasizing that this operator contains more structure than is typically made explicit in the conventional amplitude-based description. 
In particular, it encodes not only transition probabilities but also the coherent mixing between different helicity configurations. 
As such, the $T$-matrix may be viewed as defining a transformation on a finite-dimensional state space, rather than merely a collection of individual amplitudes.

This observation suggests that, beyond its role in computing cross sections, the scattering amplitude naturally admits an interpretation as a structured linear transformation acting on internal degrees of freedom. 
In the following, we will make this structure explicit by decomposing this transformation into simpler components, in the quantum circuit language.

\subsection{Decomposition into elementary operations}

Having established that the scattering process defines a finite-dimensional linear transformation on the helicity space, we now address how this transformation can be expressed in terms of simpler components.

A useful starting point is the observation that any complex matrix can be decomposed into a sequence of basis changes and a diagonal action. Concretely, for a given transformation $T$, one may write
\begin{equation}
T = U \, \Sigma \, V^\dagger,
\end{equation}
where $U$ and $V$ are unitary matrices, and $\Sigma$ is a diagonal matrix with non-negative real entries. This is known as the singular value decomposition (SVD). In this form, the action of $T$ can be understood as follows: the matrix $V^\dagger$ first performs a change of basis in the initial helicity space, the diagonal matrix $\Sigma$ assigns relative weights to different components, and $U$ then mixes these components into the final helicity configurations.

This decomposition already separates two distinct types of structure present in the scattering process. The unitary matrices $U$ and $V$ describe coherent mixing between helicity states, while the diagonal matrix $\Sigma$ encodes non-unitary effects, reflecting the fact that the restricted helicity space does not form a closed system. In particular, the singular values in $\Sigma$ determine the effective strength with which different components are transmitted through the process.

To further resolve the structure of the unitary matrices $U$ and $V$, one may decompose them into a sequence of simpler operations acting on pairs of states. A convenient way to achieve this is through the cosine-sine decomposition (CSD), which expresses a unitary matrix in terms of rotations in two-dimensional subspaces. In its simplest form, this decomposition reduces a unitary transformation into a product of elementary blocks that act non-trivially only on two-dimensional sectors, with each block taking the form of a rotation characterized by an angle.
Taking $U$ for example,
\begin{align}
U & =\begin{pmatrix}U_{1} & 0\\
0 & U_{2}
\end{pmatrix}\begin{pmatrix}C_{U} & S_{U}\\
-S_{U} & C_{U}
\end{pmatrix}\begin{pmatrix}U_{3} & 0\\
0 & U_{4}
\end{pmatrix},
\end{align}
where $U_{i}$ are $2\times2$ unitary matrices which can be further decomposed using CSD, and
the matrices $C_{U,V}=\mathrm{diag}(\cos\alpha_{i}),~S_{U,V}=\mathrm{diag}(\sin\alpha_{i})$
are diagonal $2\times2$ matrices whose diagonal entries are cosines
and sines of rotation angles.

From this perspective, the full transformation $T$ may be viewed as a sequence of elementary operations on the helicity space: basis changes that mix components, interleaved with diagonal operations that assign relative weights. This provides an operational description of the scattering process, in which the amplitude is reorganized into a structured sequence of transformations rather than a collection of individual matrix elements.

In general, such decompositions are implemented numerically. However, for the simple scattering process considered in this work, the structure of the transformation is sufficiently constrained that the decomposition can be carried out analytically. This allows the sequence of elementary operations to be written in analytical form, making the underlying structure of the process directly visible.

\subsection{Interpretation as a sequence of quantum operations}

The decomposition obtained in the previous subsection admits a natural representation in terms of elementary quantum operations. 
We now express this structure using a standard circuit language.

The basic building blocks are operations acting on two-dimensional subspaces of the helicity space. 
Such operations can be described in terms of rotations generated by Pauli matrices. 
In particular, in terms of standard gates $R_y(\theta) = e^{-i \theta \sigma_y / 2}$, and $R_z(\phi) = e^{-i \phi \sigma_z / 2}$, where $\sigma_y$ and $\sigma_z$ are Pauli matrices acting on a chosen two-dimensional sector. 
The operator $R_y(\theta)$ describes mixing between two helicity states, while $R_z(\phi)$ introduces a relative phase between them.

With these definitions, the two-dimensional rotation blocks appearing in the decomposition of unitary matrices can be directly identified with $R_y$-type operations, possibly accompanied by $R_z$ phase rotations. 
In this way, the unitary part of the scattering transformation is expressed as a sequence of elementary rotations acting on pairs of helicity states.

We now turn to the non-unitary part of the transformation, encoded in the diagonal matrix $\Sigma$,
\begin{equation}
\Sigma = \mathrm{diag}(s_1, s_2, \dots).
\end{equation}
Unlike the unitary rotations above, this operation assigns different weights to different components and therefore cannot be interpreted as a reversible transformation. 
Its effect is to attenuate or enhance specific helicity components.

This non-unitary structure has a clear physical origin. As discussed in Section~\ref{sec:IIA}, the restriction to a fixed scattering channel and helicity sector does not define a closed system. Part of the amplitude can flow into degrees of freedom that are not included in the reduced description, and the diagonal entries of $\Sigma$ encode the effective transmission strength of the components that remain.

In the language of quantum circuits, such non-unitary operations are typically associated with conditional processes, such as postselection or the restriction to a subset of outcomes. In the present context, however, they arise directly from the structure of the scattering process itself, which can be naturally formulated within the framework of quantum operations. In general, a quantum operation acting on a
density matrix $\rho$ can be written as~\citep{Kraus1983,Choi1975} 

\begin{equation}
\mathcal{E}(\rho)=\sum_{k}K_{k}\rho K_{k}^{\dagger}=K_{0}\rho K_{0}^{\dagger}+K_{1}\rho K_{1}^{\dagger}+\ldots,
\end{equation}
where the Kraus operators $K_{k}$ satisfy $\sum_{k}K_{k}^{\dagger}K_{k}\leq I$.
Each operator $K_{k}$ corresponds to a possible measurement outcome,
with the full set encoding all scattering channels. 

The non-unitary matrix $\Sigma$ can be interpreted as the Kraus operator $K_{0}$ corresponding to the selected scattering channel, while a complementary operator $K_{1}$ accounts for all other unobserved scattering outcomes
\begin{equation}
K_{0}=\Sigma,  \quad   K_{1}=\sqrt{I-\Sigma^{\dagger}\Sigma}.
\end{equation}

Combining these elements, the scattering process can be represented as a sequence of elementary operations acting on the helicity space: rotations of the $R_y$ and $R_z$ type, describing coherent mixing and phase evolution, interleaved with a non-unitary diagonal operation $K_{0}$ that encodes the effective loss of information
\begin{equation}
|\alpha\rangle\rightarrow  U \, K_{0} \, V^\dagger |\alpha\rangle.
\end{equation}
This establishes a direct bridge between the amplitude-based description and a circuit representation of the process.

\section{\protect\label{sec:The-process}The result of $e^{+}e^{-}\to\mu^{+}\mu^{-}$}

\subsection{$T$-matrix of $e^{+}e^{-}\to\mu^{+}\mu^{-}$ and the quantum circuit }

We consider the high-energy scattering process $e^{+}e^{-}\to\mu^{+}\mu^{-}$, including contributions from both QED and the weak interaction at tree level, whose amplitude is given in Appendix~\ref{sec:Amplitude}. The helicity of each lepton, being either right-handed (R) or left-handed (L), spans a two-dimensional space that can be identified with a single qubit, with the computational basis defined as $|0\rangle \equiv |R\rangle$ and $|1\rangle \equiv |L\rangle$. Consequently, the scattering process can be viewed as a quantum circuit acting on a two-qubit system.

To facilitate comparison with realistic quantum devices, we assume that the polarization fractions of the final state can be experimentally accessed. In the case of muons, their polarization can be inferred from the angular distributions of their decay products~\citep{Arbuzov:2023qgc,Super-Kamiokande:2024rwz}. For the initial state, realistic collider experiments yield partially polarized beams. Such imperfections may be described by a density operator rather than a pure state vector. However, these considerations do not modify the circuit-level structure of the mapping we construct. Moreover, certain algorithms, such as trace-evaluation protocols with $\sum_n \langle n|U|n\rangle = \mathrm{Tr}(U)$, can be implemented using mixed-state inputs without requiring coherent state preparation~\citep{Knill:1998wi}.

Ordering
the basis as $(RR,RL,LR,LL)$, we can write the $T$-matrix in the helicity amplitudes
\begin{equation}
T=\begin{pmatrix}\mathcal{M}_{RR\to RR} & \mathcal{M}_{RR\to RL} & \mathcal{M}_{RR\to LR} & \mathcal{M}_{RR\to LL}\\
\mathcal{M}_{RL\to RR} & \mathcal{M}_{RL\to RL} & \mathcal{M}_{RL\to LR} & \mathcal{M}_{RL\to LL}\\
\mathcal{M}_{LR\to RR} & \mathcal{M}_{LR\to RL} & \mathcal{M}_{LR\to LR} & \mathcal{M}_{LR\to LL}\\
\mathcal{M}_{LL\to RR} & \mathcal{M}_{LL\to RL} & \mathcal{M}_{LL\to LR} & \mathcal{M}_{LL\to LL}
\end{pmatrix}.
\end{equation}
Given that $m_e, m_\mu \ll m_Z$, we can neglect the lepton masses
in the high-energy regime. After this approximation, the analytical expression for the $T$-matrix is obtained
\begin{equation}
T=\begin{pmatrix}0 & 0 & 0 & 0\\
0 & a(1+\cos\theta) & b(1-\cos\theta) & 0\\
0 & b(1-\cos\theta) & d(1+\cos\theta) & 0\\
0 & 0 & 0 & 0
\end{pmatrix},\label{eq:Tmatrix}
\end{equation}
in which  $\theta$ is the scattering angle in the center-of-mass frame, and $a,b$ and $d$ are auxiliary functions given by

\begin{align}
a & =\left(\frac{4\mu^{2}g_{R}^{2}}{4\mu^{2}-1}\sec^{2}\theta_{W}+\sin^{2}\theta_{W}\right),\label{eq:a-function}\\
b & =\left(\frac{4\mu^{2}g_{R}g_{L}}{4\mu^{2}-1}\sec^{2}\theta_{W}+\sin^{2}\theta_{W}\right),\label{eq:b-function}\\
d & =\left(\frac{4\mu^{2}g_{L}^{2}}{4\mu^{2}-1}\sec^{2}\theta_{W}+\sin^{2}\theta_{W}\right),\label{eq:d-function}
\end{align}
in which $\theta_{W}$ is the Weinberg angle, and $g_L$ and $g_R$ denote the left- and right-handed couplings of the leptons to the $Z$ boson.
 In the above expressions, we have defined
a dimensionless parameter $\mu\equiv|\vec{p}|/m_{Z}$, where $\vec{p}$
is the 3-momentum of the incoming electron.

\begin{figure}[tb]
\centering
\includegraphics[scale=0.85]{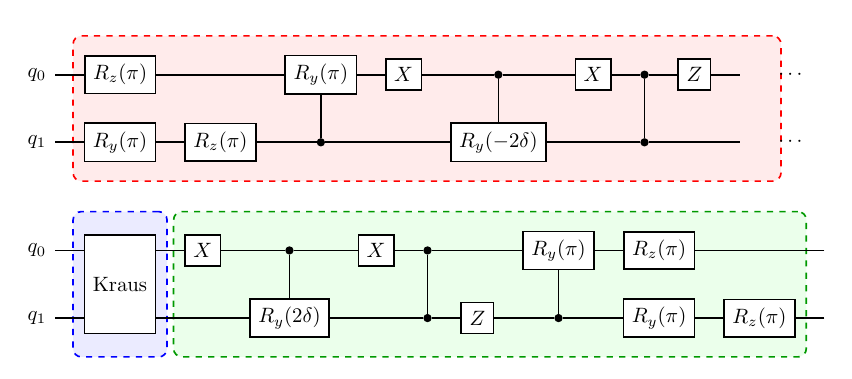}

\caption{\protect\label{fig:full-circuit}Quantum circuit representation of
the helicity transition matrix $T$ for the collider process $e^{-}e^{+}\to\mu^{-}\mu^{+}$.
The circuit implements the analytical decomposition $T=U\Sigma V^{\dagger}$,
where the unitary components $U$ (red block) and $V^{\dagger}$ (green
block) are realized by sequences of single-qubit and two-qubit gates,
and the non-unitary diagonal matrix $\Sigma$ (purple block) is
interpreted as a Kraus operator associated with the postselected channel.
The qubits $q_{0}$ and $q_{1}$ encode the helicity degrees of freedom. Here $X$, $Y$, and $Z$ denote the Pauli gates acting on a single qubit, and controlled-$R_y$ gates describe mixing between two qubits, in which $\delta$ depends on the scattering angle $\theta$ at a fixed collider energy.}
\end{figure}

We perform the SVD of the matrix in Eq.~(\ref{eq:Tmatrix}), and the analytical results are given in Appendix~\ref{sec:Analytical-results-of-SVD}. The CSD
of the resulting unitary matrices, and their quantum circuit
realizations, are given in Appendix~\ref{sec:Analytical-results-of-CSD}.
The non-unitary matrix is realized by Kraus operators. By assembling the unitary and  non-unitary components, we obtain the complete circuit, as shown in Figure~\ref{fig:full-circuit}.  This is a two-qubit quantum circuit, where the wires $q_0$ and $q_1$ encode the helicity degrees of freedom of the leptons. $X$, $Y$, and $Z$ denote the Pauli gates acting on a single qubit, and  controlled-$R_y(\pm 2 \delta)$ gates describe mixing between two qubits, in which $\delta$ depends on the scattering angle $\theta$ at a fixed collider energy. The numerical values of the electroweak parameters and auxiliary functions are given in Table~\ref{tab:parameter}.

\begin{table}[tbph]
\centering
\begin{tabular}{c c | c c}
\hline 
parameter & value & function & value\tabularnewline
\hline 
$\mu$ & 10.989 & $a(\mu,\sin^{2}\theta_{W},g_{L},g_{R})$ & 0.301\tabularnewline
$\sin^{2}\theta_{W}$ & 0.231 & $b(\mu,\sin^{2}\theta_{W},g_{L},g_{R})$ & 0.219\tabularnewline
$g_{L}$ & -0.038 & $d(\mu,\sin^{2}\theta_{W},g_{L},g_{R})$ & 0.232\tabularnewline
$g_{R}$ & 0.231 &  & \tabularnewline
\hline 
\end{tabular}

\caption{\protect\label{tab:parameter}The numerical values of the electroweak parameters  and auxiliary functions used in this work. The electroweak parameter values  are taken from the Particle Data Group~\citep{ParticleDataGroup:2024cfk}. The dimensionless parameter $\mu$ is
defined as $\mu\equiv|\vec{p}|/m_{Z}$. In this work we set the collider energy at $|\vec{p}|=1~\mathrm{TeV}$.  Auxiliary functions $a$, $b$, and $d$ depend on $\mu$, $\sin^{2}\theta_{W}$,
$g_{L}$ and $g_{R}$, and their analytical expressions are given in Eqs.~(\ref{eq:a-function})--(\ref{eq:d-function}).}
\end{table}

\subsection{\label{sec:properties}Properties of the quantum circuit}

The structure of the quantum circuit derived from the helicity transition matrix can be understood directly in terms of the underlying symmetries and kinematics of the scattering process. In particular, several nontrivial features of the circuit follow as necessary consequences of the chiral structure of the interaction, rather than as accidental properties of a specific decomposition.

We first consider the dimensionality of the effective dynamics. In the massless limit, helicity coincides with chirality for fermions, and the electroweak interaction couples left- and right-handed components in a chiral manner. As a result, scattering amplitudes with identical initial helicities vanish, and the transition matrix has support only on the subspace spanned by the $RL$ and $LR$ states. It follows that the helicity transition matrix has rank at most two. This rank reduction is therefore not a numerical property of the matrix, but a direct consequence of chiral symmetry.

\begin{figure}[tb]
\centering
\includegraphics[scale=0.4]{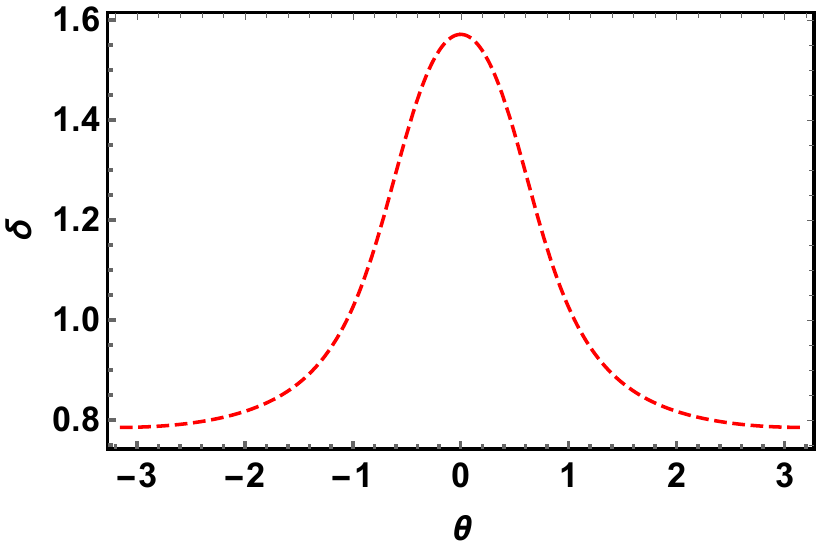}\includegraphics[scale=0.4]{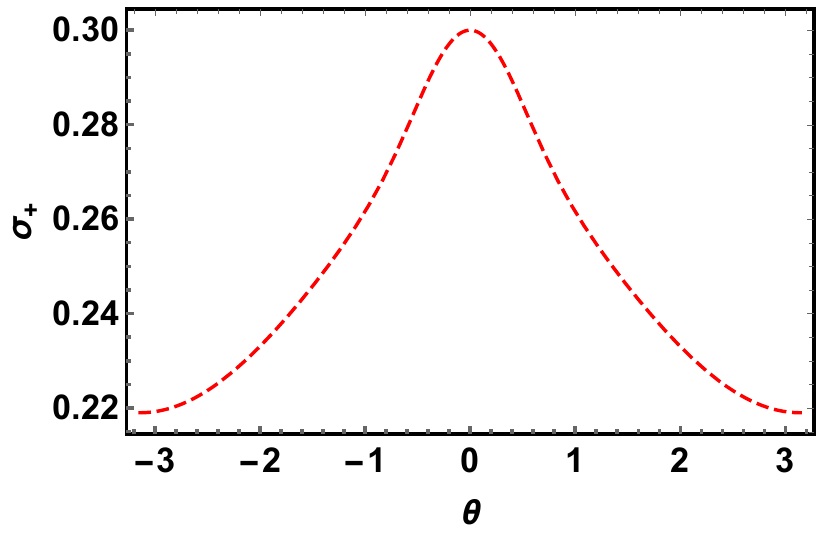}\includegraphics[scale=0.4]{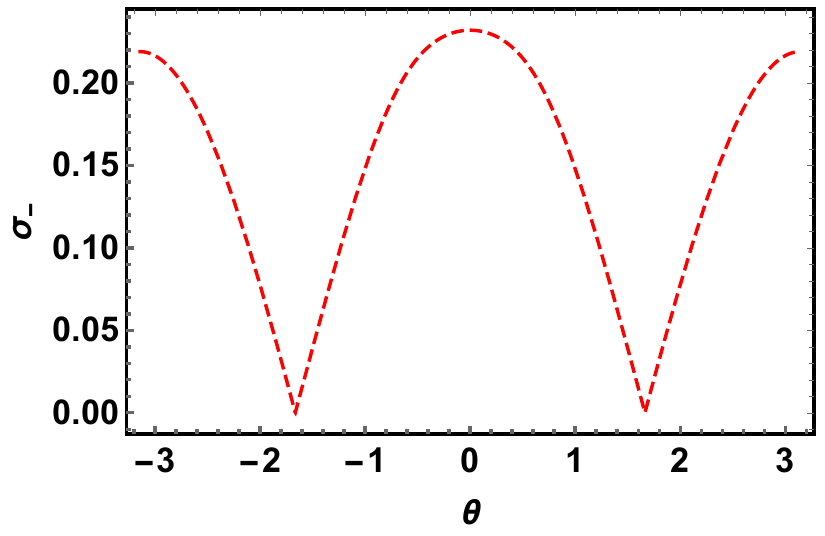}

\caption{\protect\label{fig:plots}Dependence of the quantum circuit parameters
on the scattering angle $\theta$. Left: the rotation angle $\delta(\theta)$.
Middle: the singular value $\sigma_{+}(\theta)$. Right: the singular
value $\sigma_{-}(\theta)$.}
\end{figure}

As a consequence, the nontrivial part of the scattering process is completely captured by an effective $2\times 2$ transformation acting on the $(RL,LR)$ subspace. In the circuit representation, this implies that the full process is equivalent to a single effective two-level system, whose dynamics can be described by a minimal set of parameters. In particular, the decomposition involves one mixing angle $\delta(\theta)$ arising from the unitary part, and two singular values $\sigma_{+}(\theta)$ and $\sigma_{-}(\theta)$ associated with the non-unitary component. The analytical forms of the functions $\delta(\theta)$ and $\sigma_{\pm}(\theta)$ are given in Appendices~\ref{sec:Analytical-results-of-SVD} and \ref{sec:Analytical-results-of-CSD} (see Eqs.~(\ref{eq:delta}), (\ref{eq:sigma+}), and (\ref{eq:sigma-})), and their curves are shown in Figure~\ref{fig:plots}. This reduction demonstrates that the apparent four-dimensional helicity space contains only a single dynamically active degree of freedom.

Within this reduced description, the unitary and non-unitary components play distinct roles. The rotation angle $\delta(\theta)$ encodes the coherent mixing between the $RL$ and $LR$ configurations, and is determined by the interference structure of the scattering amplitudes. In contrast, the singular values $\sigma_{+}(\theta)$ and $\sigma_{-}(\theta)$ quantify the effective transmission strength of different components after restriction to the selected channel. These quantities are not independent: both arise from the same underlying amplitudes and are therefore constrained by the scattering dynamics.

\begin{figure}
\centering
\includegraphics[scale=0.5]{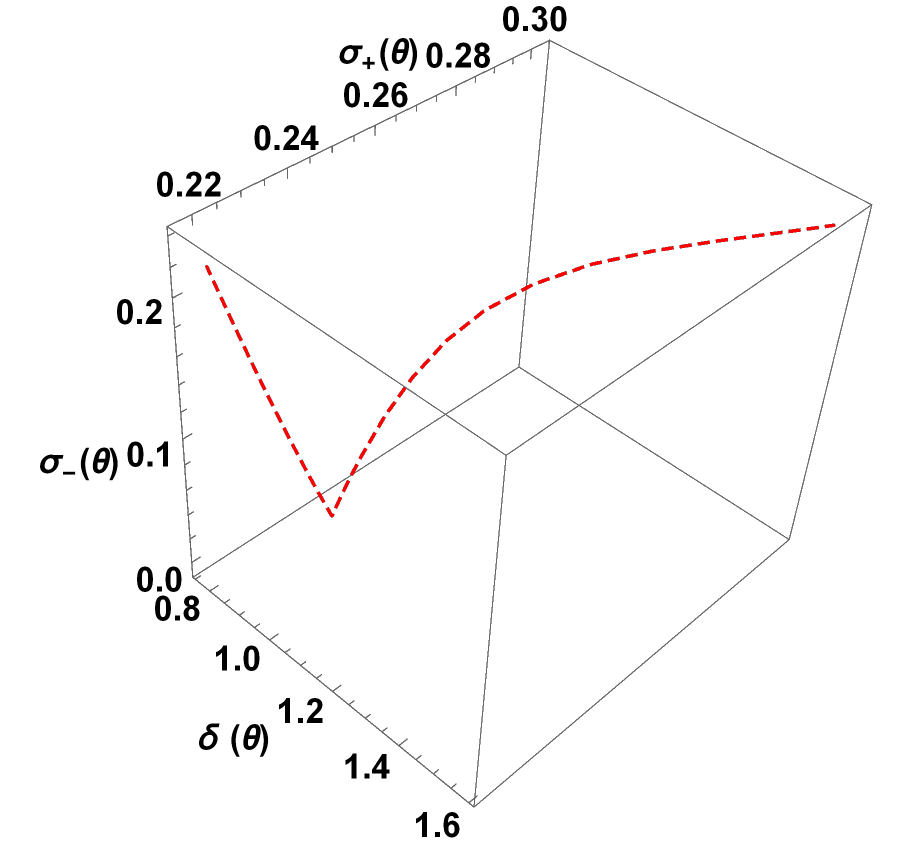}

\caption{\protect\label{fig:Trajectory3D}Trajectory of the circuit parameters
$(\delta(\theta),\sigma_{+}(\theta),\sigma_{-}(\theta))$ as the scattering
angle $\theta$ varies. The curve illustrates that the unitary and
non-unitary components of the decomposition are dynamically correlated
through a one-dimensional curve in parameter space.}

\end{figure}

This constraint becomes particularly transparent when the circuit parameters are viewed as functions of the scattering angle $\theta$. Although the decomposition formally introduces three parameters $(\delta,\sigma_{+},\sigma_{-})$, they do not span a three-dimensional parameter space. Instead, as $\theta$ varies, the system traces out a one-dimensional trajectory in this space, as shown in Figure~\ref{fig:Trajectory3D}. This demonstrates that the unitary and non-unitary structures are not freely adjustable, but are correlated through the kinematic dependence of the scattering process.

A particularly striking manifestation of this correlation occurs at scattering angles where one of the singular values approaches zero. At these points, the transformation effectively reduces to a rank-one map, indicating that only a single helicity channel contributes significantly to the process. Physically, this corresponds to destructive interference suppressing one component of the amplitude. In the circuit language, this appears as a collapse of the non-unitary operation to an effectively one-dimensional projection, making the loss of information explicit.

Taken together, these results show that the circuit representation does more than provide an alternative description of the scattering amplitude. It reorganizes the dynamics into a set of constrained operations, in which symmetry and kinematics directly determine the allowed structure of both the unitary mixing and the non-unitary attenuation. In this sense, the properties of the circuit can be viewed as a direct manifestation of the underlying physical principles governing the scattering process.

\subsection{Advantages of the circuit representation}

The circuit representation introduced above does not change the physical content of the scattering process, but reorganizes it into a form that makes certain structures more directly accessible. This can be seen explicitly in the example discussed in this section.

In the amplitude-based description, the information about the process is encoded in the pattern of matrix elements of the helicity transition matrix. While this representation is complete, different physical effects--such as helicity mixing, relative phases, and the effective suppression of certain channels--are intertwined within these elements. Extracting these features typically requires combining several entries and identifying the relevant interference patterns.

In contrast, the circuit representation separates these effects at the level of the description. The decomposition organizes the transformation into a sequence of elementary operations, in which coherent mixing, phase evolution, and non-unitary attenuation appear as distinct components. In particular, the non-unitary structure, which originates from the restriction to a reduced helicity sector, is made explicit as a well-defined operation, rather than being implicitly encoded in the amplitudes.

This reorganization also highlights constraints that are not immediately transparent in the amplitude form. As shown in Section~\ref{sec:properties}, the parameters describing the circuit are not independent, but are restricted to a one-dimensional manifold determined by the scattering kinematics. In the circuit language, this appears naturally as a correlation between the unitary and non-unitary parts of the transformation.

More generally, the circuit representation emphasizes the process-like nature of the scattering dynamics. Instead of presenting only the net transition between initial and final states, it provides a structured description in terms of intermediate operations acting on the internal degrees of freedom. This perspective offers a complementary way to analyze scattering processes, in which symmetry and kinematics are directly reflected in the organization of the transformation.

\section{Summary}

In this work, we have shown that a scattering process, once restricted to a fixed channel and helicity sector, admits a natural description as a sequence of elementary operations acting on a finite-dimensional space of internal degrees of freedom. Rather than viewing the amplitude solely as a collection of matrix elements, this perspective organizes the process as a structured transformation, in which coherent mixing, phase evolution, and non-unitary attenuation appear as distinct components.

This structure is not imposed, but follows directly from the properties of the helicity transition matrix. The decomposition of the transformation makes explicit the interplay between symmetry and kinematics, which constrains both the unitary and non-unitary parts of the process. In particular, as demonstrated in the example considered here, the apparent degrees of freedom in the circuit representation are highly restricted, leading to correlated behavior that reflects the underlying dynamics.

The circuit representation therefore provides a complementary way to analyze scattering processes, in which the focus shifts from individual amplitudes to the organization of the transformation itself. This viewpoint emphasizes the process-like nature of the dynamics and makes structural features directly accessible, offering a useful framework for exploring the quantum information content of high-energy processes.

\section*{Acknowledgments}

WX was supported in part by the National Natural Science Foundation
of China under Grants No. 12375137 and 12005114. J.-C. Y. was supported
in part by the National Natural Science Foundation of China under
Grants Nos. 11875157 and 12147214, and the Natural Science Foundation of the Liaoning Scientific Committee No.~LJKMZ20221431.

\appendix

\section{\protect\label{sec:Amplitude}Scattering amplitude of $e^{+}e^{-}\to\mu^{+}\mu^{-}$}

\begin{figure}
\centering
\includegraphics{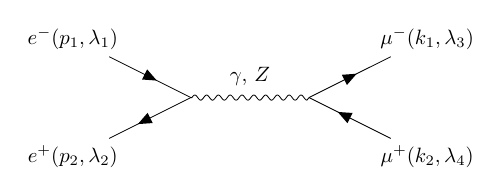}

\caption{\protect\label{fig:Feynman-diagram}Tree-level Feynman diagram for
the process $e^{-}(p_{1},\lambda_{1})+e^{+}(p_{2},\lambda_{2})\to\mu^{-}(k_{1},\lambda_{3})+\mu^{+}(k_{2},\lambda_{4})$,
mediated by photon and $Z$-boson exchange. The four-momenta of the
incoming electron and positron are denoted by $p_{1}$ and $p_{2}$,
respectively, while those of the outgoing muon and antimuon are $k_{1}$
and $k_{2}$. The labels $\lambda_{1,2}$ and $\lambda_{3,4}$ denote
the helicities of the initial and final state fermions, respectively.}
\end{figure}

 At tree level, the process
$e^{+}e^{-}\to\mu^{+}\mu^{-}$ receives contributions from two $s$-channel
diagrams: photon exchange and $Z$-boson exchange. The electron-photon
and muon-photon vertices are $-ie\gamma^{\mu}$, and the electron(muon)-$Z$
vertex is 
\begin{equation}
-i\frac{g}{\cos\theta_{W}}\gamma^{\mu}\,\frac{1}{2}\left(c_{V}^{f}-c_{A}^{f}\gamma_{5}\right),
\end{equation}
with 
\begin{equation}
c_{V}^{f}=-\frac{1}{2}+2\sin^{2}\theta_{W},\qquad c_{A}^{f}=-\frac{1}{2}
\end{equation}
for both electrons and muons. $g$ is the weak coupling constant and
$\theta_{W}$ is the Weinberg angle. The tree-level $s$-channel diagram
for the process $e^{+}e^{-}\to\mu^{+}\mu^{-}$ is shown in Figure~\ref{fig:Feynman-diagram}.
The scattering amplitude including both photon and $Z$ boson exchange
contributions is given by
\begin{align}
i\mathcal{M} & =\bar{v}(p_{2},\lambda_{2})(ie\gamma^{\mu})u(p_{1},\lambda_{1})\,\frac{-ig_{\mu\nu}}{q^{2}}\,\bar{u}(k_{1},\lambda_{3})(ie\gamma^{\nu})v(k_{2},\lambda_{4})\\
 & +\;\bar{v}(p_{2},\lambda_{2})\left(i\frac{g}{c_{W}}\gamma^{\mu}(g_{L}P_{L}+g_{R}P_{R})\right)u(p_{1},\lambda_{1})\,\frac{-ig_{\mu\nu}}{q^{2}-m_{Z}^{2}}\,\bar{u}(k_{1},\lambda_{3})\left(i\frac{g}{c_{W}}\gamma^{\nu}(g_{L}P_{L}+g_{R}P_{R})\right)v(k_{2},\lambda_{4}),
\end{align}
in which $q=p_{1}+p_{2}=k_{1}+k_{2}$, $s_{W}=\sin\theta_{W}$ and
$c_{W}=\cos\theta_{W}$. The coupling constants satisfy the relation
$g\sin\theta_{W}=e$. The chiral projection operators are $P_{L,R}=\frac{1\mp\gamma^{5}}{2}.$
In the above equation we have defined 
\begin{equation}
\begin{alignedat}{1}g_{L} & =\frac{c_{V}+c_{A}}{2}=-\frac{1}{2}+\sin^{2}\theta_{W},\\
g_{R} & =\frac{c_{V}-c_{A}}{2}=\sin^{2}\theta_{W}.
\end{alignedat}
\end{equation}

\section{\protect\label{sec:Analytical-results-of-SVD}Analytical results
of SVD}

The singular value decomposition (SVD) provides a general representation
of any finite-dimensional matrix in terms of unitary transformations
and a diagonal scaling. Specifically, an arbitrary matrix $T$ can
be written as $T=U\Sigma V^{\dagger}$, where $U$ and $V$ are unitary
matrices and $\Sigma$ is a diagonal matrix with non-negative entries,
known as the singular values. This decomposition separates $T$ into
both its unitary and non-unitary components. For the $T$-matrix given
in Eq.~(\ref{eq:Tmatrix}), we obtain the analytical results of $U$,
$V^{\dagger}$ and $\Sigma$.

The expression of $U$ is given by

\begin{equation}
U=\left(\begin{array}{cccc}
u_{11} & u_{12} & u_{13} & u_{14}\\
u_{21} & u_{22} & u_{23} & u_{24}\\
u_{31} & u_{32} & u_{33} & u_{34}\\
u_{41} & u_{42} & u_{43} & u_{44}
\end{array}\right),
\end{equation}
with $u_{11}=u_{12}=u_{13}=u_{23}=u_{24}=u_{33}=u_{34}=u_{41}=u_{42}=u_{44}=0$,
$u_{14}=u_{43}=1$ and {\scriptsize
\begin{align}
u_{21} & =\frac{4\left(\cos(\theta)\left(a^{2}-ad+2b^{2}\right)+a\csc^{2}\left(\frac{\theta}{2}\right)(-2a+2d+k)+a(3a-3d-k)-2b^{2}\right)}{\sqrt{\csc^{4}\left(\frac{\theta}{2}\right)\left(\cos(\theta)\left(4a^{2}-2a(2d+k)-8b^{2}\right)+\cos(2\theta)\left(a^{2}-ad+2b^{2}\right)+a(3a-3d-2k)+6b^{2}\right)^{2}+16b^{2}((a+d)\cos(\theta)+a+d-k)^{2}}},\\
u_{22} & =\frac{4\left(\cos(\theta)\left(a^{2}-ad+2b^{2}\right)+\frac{a(\cos(\theta)(3a-3d+k)+a-d+k)}{\cos(\theta)-1}-2b^{2}\right)}{\sqrt{\csc^{4}\left(\frac{\theta}{2}\right)\left(\cos(2\theta)\left(a^{2}-ad+2b^{2}\right)+2\cos(\theta)\left(a(2a-2d+k)-4b^{2}\right)+a(3a-3d+2k)+6b^{2}\right)^{2}+16b^{2}((a+d)\cos(\theta)+a+d+k)^{2}}},\\
u_{31} & =-\frac{4b((a+d)\cos(\theta)+a+d-k)}{\sqrt{\csc^{4}\left(\frac{\theta}{2}\right)\left(\cos(\theta)\left(4a^{2}-2a(2d+k)-8b^{2}\right)+\cos(2\theta)\left(a^{2}-ad+2b^{2}\right)+a(3a-3d-2k)+6b^{2}\right)^{2}+16b^{2}((a+d)\cos(\theta)+a+d-k)^{2}}},\\
u_{32} & =-\frac{4b\left(a+d+k+(a+d)\cos(\theta)\right)}{\sqrt{16b^{2}\left(a+d+k+(a+d)\cos(\theta)\right)^{2}+\left(6b^{2}+a(3a-3d+2k)+2\left(-4b^{2}+a(2a-2d+k)\right)\cos(\theta)+(a^{2}+2b^{2}-ad)\cos(2\theta)\right)^{2}\csc^{4}\left(\frac{\theta}{2}\right)}}.
\end{align}
}In the above expressions $a,b$ and $d$ are functions defined in
Eqs.~(\ref{eq:a-function})--(\ref{eq:d-function}), and $k$ is an
auxiliary function, which itself is a function of $\theta$ and $a,b,d$
\begin{equation}
k=\sqrt{2\cos\theta\left[(a-d)^{2}-4b^{2}\right]+\frac{1}{2}\left[\cos(2\theta)+3\right]\left[(a-d)^{2}+4b^{2}\right]}.
\end{equation}

The expression of $V^{\dagger}$ is given by
\begin{equation}
V^{\dagger}=\left(\begin{array}{cccc}
v_{11} & v_{12} & v_{13} & v_{14}\\
v_{21} & v_{22} & v_{23} & v_{24}\\
v_{31} & v_{32} & v_{33} & v_{34}\\
v_{41} & v_{42} & v_{43} & v_{44}
\end{array}\right),
\end{equation}
with $v_{11}=v_{14}=v_{21}=v_{24}=v_{31}=v_{32}=v_{33}=v_{42}=v_{43}=v_{44}=0$,
$v_{34}=v_{41}=1$ and{\scriptsize
\begin{align}
v_{12} & =\frac{\cos^{2}\left(\frac{\theta}{2}\right)\csc^{2}(\theta)((a-d)\cos(\theta)+a-d-k)}{b\sqrt{\frac{\csc^{4}\left(\frac{\theta}{2}\right)((d-a)\cos(\theta)-a+d+k)^{2}}{16b^{2}}+1}},\\
v_{13} & =\frac{1}{\sqrt{\frac{\csc^{4}\left(\frac{\theta}{2}\right)((d-a)\cos(\theta)-a+d+k)^{2}}{16b^{2}}+1}},\\
v_{22} & =\frac{\cos^{2}\left(\frac{\theta}{2}\right)\csc^{2}(\theta)((a-d)\cos(\theta)+a-d+k)}{b\sqrt{\frac{\csc^{4}\left(\frac{\theta}{2}\right)((a-d)\cos(\theta)+a-d+k)^{2}}{16b^{2}}+1}},\\
v_{23} & =\frac{1}{\sqrt{\frac{\csc^{4}\left(\frac{\theta}{2}\right)((a-d)\cos(\theta)+a-d+k)^{2}}{16b^{2}}+1}}.
\end{align}
}{\scriptsize\par}

The expression of $\Sigma$ is given by
\begin{equation}
\Sigma=\begin{pmatrix}\sigma_{+} & 0 & 0 & 0\\
0 & \sigma_{-} & 0 & 0\\
0 & 0 & 0 & 0\\
0 & 0 & 0 & 0
\end{pmatrix},
\end{equation}
with {\scriptsize
\begin{align}
\sigma_{+} & =\frac{1}{4}\sqrt{8(a^{2}+d^{2})\cos^{4}\left(\frac{\theta}{2}\right)+2\sqrt{2}\,\sqrt{(a+d)^{2}\cos^{4}\left(\frac{\theta}{2}\right)-\left(4(-4b^{2}+(a-d)^{2})\cos(\theta)+(4b^{2}+(a-d)^{2})(3+\cos(2\theta))\right)}+16b^{2}\sin\left(\frac{\theta}{2}\right)^{4}},\label{eq:sigma+}\\
\sigma_{-} & =\frac{1}{4}\sqrt{8(a^{2}+d^{2})\cos^{4}\left(\frac{\theta}{2}\right)-2\sqrt{2}\,\sqrt{(a+d)^{2}\cos^{4}\left(\frac{\theta}{2}\right)-\left(4(-4b^{2}+(a-d)^{2})\cos(\theta)+(4b^{2}+(a-d)^{2})(3+\cos(2\theta))\right)}+16b^{2}\sin\left(\frac{\theta}{2}\right)^{4}}.\label{eq:sigma-}
\end{align}
}{\scriptsize\par}

\section{\protect\label{sec:Analytical-results-of-CSD}Analytical results
of CSD}

The cosine-sine decomposition (CSD) expresses a unitary matrix as
a sequence of block-diagonal unitaries and rotation blocks. This form
is particularly useful for constructing quantum circuit implementations.

The explicit form of the CSD of $U$ matrix is given by

\begin{align}
 & \begin{pmatrix}U_{1} & 0\\
0 & U_{2}
\end{pmatrix}=\left(\begin{array}{cc|cc}
0 & 1 & 0 & 0\\
1 & 0 & 0 & 0\\
\hline 0 & 0 & 0 & -1\\
0 & 0 & -1 & 0
\end{array}\right),\\
 & \begin{pmatrix}C_{U} & S_{U}\\
-S_{U} & C_{U}
\end{pmatrix}=\left(\begin{array}{cc|cc}
1 & 0 & 0 & 0\\
0 & 0 & 0 & -1\\
\hline 0 & 0 & 1 & 0\\
0 & 1 & 0 & 0
\end{array}\right),\\
 & \begin{pmatrix}U_{3} & 0\\
0 & U_{4}
\end{pmatrix}=\left(\begin{array}{cc|cc}
-\cos\delta & \sin\delta & 0 & 0\\
\sin\delta & \cos\delta & 0 & 0\\
\hline 0 & 0 & -1 & 0\\
0 & 0 & 0 & -1
\end{array}\right).
\end{align}
In our construction, the gate parameter $\delta$ is determined by
the underlying scattering kinematics and is explicitly expressed as
a function of the scattering angle $\theta$
\begin{equation}
\cos\delta=\frac{1}{\sqrt{\frac{\left[(a-d)\cos\theta+a-d+k\right]{}^{2}\left(\frac{\csc^{4}\frac{\theta}{2}\left[(d-a)\cos\theta-a+d+k\right]{}^{2}}{16b^{2}}+1\right)}{\left[(d-a)\cos\theta-a+d+k\right]{}^{2}\left(\frac{\csc^{4}\frac{\theta}{2}\left[(a-d)\cos\theta+a-d+k\right]{}^{2}}{16b^{2}}+1\right)}+1}},\label{eq:delta}
\end{equation}

After performing the CSD, the unitary matrix $U$ is expressed as
a product of block-diagonal unitary matrices and a central cosine-sine
rotation block. For the two-qubit system considered here, these matrices
act on a four-dimensional Hilbert space and can therefore be directly
mapped to quantum gates acting on two qubits. The block-diagonal unitary
matrices correspond to controlled single-qubit rotations, and the
central cosine-sine block can be realized using controlled rotation
gates. Consequently, the entire unitary operator $U$ can be compiled
into a sequence of elementary two-qubit operations, forming an explicit
quantum circuit acting on the two-qubit helicity space
\begin{align}
 & \begin{pmatrix}U_{1} & 0\\
0 & U_{2}
\end{pmatrix}=\left[e^{i\pi}\left(\begin{array}{cc}
1 & 0\\
0 & e^{i\pi}
\end{array}\right)\otimes\left(\begin{array}{cc}
1 & 0\\
0 & e^{i\pi}
\end{array}\right)\right]\left[\left(\begin{array}{cc}
1 & 0\\
0 & 1
\end{array}\right)\otimes\left(\begin{array}{cc}
\cos\frac{\pi}{2} & -\sin\frac{\pi}{2}\\
\sin\frac{\pi}{2} & \cos\frac{\pi}{2}
\end{array}\right)\right]=e^{i\pi}\,(Z\otimes Z)\,(I\otimes R_{y}(\pi)),\\
 & \begin{pmatrix}C_{U} & S_{U}\\
-S_{U} & C_{U}
\end{pmatrix}=\left(\begin{array}{cccc}
1 & 0 & 0 & 0\\
0 & \cos\frac{\pi}{2} & 0 & -\sin\frac{\pi}{2}\\
0 & 0 & 1 & 0\\
0 & \sin\frac{\pi}{2} & 0 & \cos\frac{\pi}{2}
\end{array}\right)=CR_{y}^{(q_{1}\to q_{0})}(\pi),\\
 & \begin{pmatrix}U_{3} & 0\\
0 & U_{4}
\end{pmatrix}=\left[e^{i\pi}\left(\begin{array}{cc|cc}
1 & 0 & 0 & 0\\
0 & e^{i\pi} & 0 & 0\\
\hline 0 & 0 & 1 & 0\\
0 & 0 & 0 & 1
\end{array}\right)\right]\left(\begin{array}{cc|cc}
\cos\delta & -\sin\delta & 0 & 0\\
\sin\delta & \cos\delta & 0 & 0\\
\hline 0 & 0 & 1 & 0\\
0 & 0 & 0 & 1
\end{array}\right)=e^{i\pi}\,(I\otimes Z)\,\text{CNOT}_{0\to1}\,(X\otimes I)\,CR_{y}(2\delta)\,(X\otimes I).
\end{align}
Here $X$, $Y$, and $Z$ denote the Pauli gates acting on a single
qubit, and $CR$ denotes a controlled rotation gate applied to the
target qubit conditioned on the control qubit being in state $|1\rangle$. 

The explicit form of the CSD of $V^{\dagger}$ matrix is given by
\begin{align}
 & \begin{pmatrix}V_{1} & 0\\
0 & V_{2}
\end{pmatrix}=\left(\begin{array}{cc|cc}
\cos\delta & \sin\delta & 0 & 0\\
-\sin\delta & \cos\delta & 0 & 0\\
\hline 0 & 0 & -1 & 0\\
0 & 0 & 0 & 1
\end{array}\right),\\
 & \begin{pmatrix}C_{V} & S_{V}\\
-S_{V} & C_{V}
\end{pmatrix}=\left(\begin{array}{cc|cc}
1 & 0 & 0 & 0\\
0 & 0 & 0 & -1\\
\hline 0 & 0 & 1 & 0\\
0 & 1 & 0 & 0
\end{array}\right),\\
 & \begin{pmatrix}V_{3} & 0\\
0 & V_{4}
\end{pmatrix}=\left(\begin{array}{cc|cc}
0 & 1 & 0 & 0\\
1 & 0 & 0 & 0\\
\hline 0 & 0 & 0 & -1\\
0 & 0 & -1 & 0
\end{array}\right),
\end{align}
with $\cos\delta$ defined in Eq.~(\ref{eq:delta}).

The unitary matrix $V^{\dagger}$ can be decomposed and implemented
in an analogous manner. Applying the CSD, $V^{\dagger}$ is expressed
as a sequence of block-diagonal unitary matrices and a central rotation
block, which can be mapped directly onto quantum gates acting on two
qubits. As in the case of $U$, the block-diagonal components correspond
to controlled single-qubit operations, while the cosine-sine block
is realized through controlled rotations. This yields a compact quantum
circuit representation of $V^{\dagger}$ acting on the two-qubit helicity
space
\begin{align}
 & \begin{pmatrix}V_{1} & 0\\
0 & V_{2}
\end{pmatrix}=\left(\begin{array}{cccc}
1 & 0 & 0 & 0\\
0 & 1 & 0 & 0\\
0 & 0 & e^{i\pi} & 0\\
0 & 0 & 0 & 1
\end{array}\right)\left(\begin{array}{cc|cc}
\cos(-\delta) & -\sin(-\delta) & 0 & 0\\
\sin(-\delta) & \cos(-\delta) & 0 & 0\\
\hline 0 & 0 & 1 & 0\\
0 & 0 & 0 & 1
\end{array}\right)=(Z\otimes I)\;\mathrm{CNOT}_{0\to1}\;(X\otimes I)\;CR_{y}(-2\delta)\;(X\otimes I),\\
 & \begin{pmatrix}C_{V} & S_{V}\\
-S_{V} & C_{V}
\end{pmatrix}=\left(\begin{array}{cccc}
1 & 0 & 0 & 0\\
0 & \cos\frac{\pi}{2} & 0 & -\sin\frac{\pi}{2}\\
0 & 0 & 1 & 0\\
0 & \sin\frac{\pi}{2} & 0 & \cos\frac{\pi}{2}
\end{array}\right)=CR_{y}^{(q_{1}\to q_{0})}(\pi),\\
 & \begin{pmatrix}V_{3} & 0\\
0 & V_{4}
\end{pmatrix}=\left[e^{i\pi}\left(\begin{array}{cc}
1 & 0\\
0 & e^{i\pi}
\end{array}\right)\otimes\left(\begin{array}{cc}
1 & 0\\
0 & e^{i\pi}
\end{array}\right)\right]\left[\left(\begin{array}{cc}
1 & 0\\
0 & 1
\end{array}\right)\otimes\left(\begin{array}{cc}
\cos\frac{\pi}{2} & -\sin\frac{\pi}{2}\\
\sin\frac{\pi}{2} & \cos\frac{\pi}{2}
\end{array}\right)\right]=e^{i\pi}\,(Z\otimes Z)\,(I\otimes R_{y}(\pi)).
\end{align}

\bibliographystyle{apsrev4-1}
\bibliography{circuit}

\end{document}